\documentclass[aps,prb,twocolumn,superscriptaddress,showpacs]{revtex4}

\usepackage{hyperref}
\usepackage[dvips]{graphicx}
\usepackage{amssymb,amsmath,amsfonts}
\usepackage[T1]{fontenc}
\usepackage{color}
\newcommand{\alfa}{$\alpha_{\mathrm{hex}}$}

\begin{document}


\title{Domains of doping in graphene on polycrystalline gold: first principles and scanning tunneling spectroscopy studies}


\author{J. S\l awi\'{n}ska}
\affiliation{Department of Theoretical Physics and Computer Science, University of Lodz, Pomorska 149/153, 90-236 Lodz, Poland}
\affiliation{Solid State Physics Department, University of Lodz, Pomorska 149/153, 90-236 Lodz, Poland}
\author{I. Wlasny}
\affiliation{Solid State Physics Department, University of Lodz, Pomorska 149/153, 90-236 Lodz, Poland}
\author{P. Dabrowski}
\affiliation{Solid State Physics Department, University of Lodz, Pomorska 149/153, 90-236 Lodz, Poland}
\author{Z. Klusek}
\affiliation{Solid State Physics Department, University of Lodz, Pomorska 149/153, 90-236 Lodz, Poland}
\author{I. Zasada}
\affiliation{Solid State Physics Department, University of Lodz, Pomorska 149/153, 90-236 Lodz, Poland}


\vspace{.15in}

\begin{abstract}
We have studied graphene/gold interface by means of density functional theory (DFT) and scanning tunneling spectroscopy (STS). Weak interaction between graphene and the underlying gold surface leaves unperturbed Dirac cones in the band-structure, but they can be shifted with respect to the Fermi level of the whole system, which results in effective doping of graphene. DFT calculations revealed that the interface is extremely sensitive to the adsorption distance and to the structure of metal's surface, in particular strong variation in doping can be attributed to the specific rearrangments of substrate's atoms, such as change of the crystallographic orientation, relaxation or other modifications of the surface. On the other hand, STS experiments have shown the presence of energetic heterogeneity in terms of the changes in the local density of states (LDOS) measured at different places on the sample. Randomly repeated regions of zero-doping and p-type doping have been identified from parabolic shape characteristics and from well defined Dirac points, respectively. The doping domains of graphene on gold seem to be related to the presence of various types of the surface structure accross the sample. DFT simulations for graphene interacting with Au have shown large differences in doping induced by considered structures of substrate, in agreement with experimental findings. All these results demonstrate the possibility of engineering the electronic properties of graphene, especially tuning the doping across one flake which can be useful for applications of graphene in electronic devices.
\end{abstract}

\pacs{73.22.Pr, 71.15.Mb, 73.20.Hb}
\maketitle

\section{Introduction}
Graphene\cite{novoselov_science}, a single layer of carbon atoms arranged as those in one sheet of graphite, has shown fascinating physical properties\cite{review} that are currently being investigated in basic research and which make it a promising material to play a key role in future microelectronics, optoelectronics and sensors\cite{nanoletters, polarizer}. Realization of most of the industrial applications requires reliable methods for production of large-area graphene, which has been practically achieved via, for example, chemical vapor deposition (CVD) growth on transistion metals\cite{iryd, ruten, platyna_sutter} as well as by ethylene irradiation in the case of graphene on copper and gold substrates\cite{brihuega}. However, the interaction with polycrystalline metallic foils can strongly affect the graphene's electronic characteristics and impair the performance of potential devices. Similarly, the presence of metallic contacts which are essential in electronics also influences its fundamental properties. Due to all these facts, mechanisms of coupling between graphene and metals is now one of the leading fields in the research of graphene.

DFT calculations and experimental studies by angle-resolved photoemission spectroscopy (ARPES) have shown that graphene can bind strongly or weakly on various metallic substrates. One can distinguish two main types of interaction between the graphene sheet and the surface of a metal, i.e. chemisorption that completely eliminates the characteristic Dirac cones from the band-structure, as for example in the case of Co, Ni and Pd substrates\cite{cobalt, nikiel2, pallad}, and physisorption which leaves the linear dispersion undisturbed, as observed for Al, Ag, Cu, Au and Pt (111) surfaces. In the case of physisorption, the weak interaction causes the Fermi level shift downwards  (upwards) with respect to the Dirac point. It results in doping by the holes (electrons) transferred from a metal to graphene which becomes p-type (n-type) doped. According to most of the theoretical studies\cite{holendrzy_prl, holendrzy, interface_xu, vanin, comparative, nasza6, nasza_cu, platyna_hiszpanie}, graphene on Al, Ag and Cu is n-type doped, while interaction with Au and Pt(111) surfaces seems to cause p-type doping. Although the particular DFT calculations can lead to different values of the doping level for the same metal, it is usually assumed that one type of substrate implies definite type and level of doping and any contradictions are related to the limitations of \textit{ab initio} methods whose results depend on the choice of the exchange-correlation interaction and on the calculation strategy\cite{nasza6}.

In our recent work, we have analyzed the correlation between the adsorption geometry and electronic properties of the graphene/Au(111) interface and we have shown the influence of the geometry parameters characterizing the modeled system\cite{nasza6}. We focused mainly on the aspects related to the strategy of the calculations, i.e. it has been found that the factors such as the in-plane cell parameter and interlayer spacing in gold can impact the Fermi-level shift and even a small change of geometry may lead to a transition between p-type and n-type doping. Furthermore, the adsorption distance which has the strongest influence on doping level (see Fig.3 in Ref. \onlinecite{nasza6}) can be, in practice, set arbitrarily, since the equilibrium position of the graphene sheet with respect to Au depends directly on the choice of one from many possible exchange-correlation functionals. 

However, it should be stressed that tuning the geometry details may be realized experimentally and should then lead to predicted abrupt modifications of electronic properties. For example the presence of  steps, buckling or corrugation of the surface could induce local variations in graphene-substrate distance. In this case graphene can be suspended above certain regions and coupled in a different way to the metal. Next, the change of the substrate's in-plane lattice parameter can be achieved by depositing graphene on another crystallographic face which is characterized not only by different lateral arrangements of metal atoms but also by different size of the unit cell of the interface. Moreover, the strong relaxation of the surface can be the consequence of interaction with impurities present on the surface or may depend on details of the sample preparation process. 

All these surprising theoretical results motivated us to experimental studies of graphene/gold system in order to verify the extreme sensitivity of the doping level on the interface structure. Since the variations of doping are expected to be local, the most suitable tool for these investigations are scanning probe methods. In particular, the STS mode provides information about first derivative of the tunneling current with respect to the bias voltage which is the measure of the local density of states at every point above the surface. The electronic structure is recorded locally, thus, in contrast to ARPES technique, one can detect nonhomogeneous electronic properties at different places on the sample and identify repeated domains of doping. 

Indeed, the preliminary STS measurements performed on graphene/gold sample have shown very strong energetic heterogeneity in terms of changes in LDOS recorded at different places on the surface\cite{klusek}. Now, we have observed separate regions with completely different electronic structures as well as slight changes in LDOS within single domains. The heterogeneity of electronic properties seems to be domains of doping, but: (i) differential tunneling spectra are not sufficient to give straightforward information about doping level unless they are completed by another experimental or theoretical technique which enables identification of the Dirac points in the profiles (see Ref.\onlinecite{nasza_cu} for wider discussion), (ii) there is no evidence which particular features in the structure of surface lead to the given changes in the doping level. It seems that the domains of doping can be created spontaneously on metallic surfaces, but more knowledge about mechanisms of interaction is needed to control the doping or to create multi-doped samples for particular purposes.

In the present paper, we attempt to make a link between theoretical models studied by means of DFT method and STS results obtained for graphene/gold system. First, we have thoroughly analyzed STS profiles of graphene/Au system and determined the character and distribution of domains with different electronic properties. Secondly, we have recorded dI/dV profiles in the regions on the sample which were not covered by graphene. This provides information of the structure of the pristine Au surface that needs to be included in the theoretical models. It should be stressed that the role of DFT simulations is twofold:(i) to identify unambiguously the Dirac points in experimental STS profiles which is achieved by providing simultaneously the information about the band-structure and LDOS above the surface according to the scheme proposed in Ref. \onlinecite{nasza_cu}, (ii) to show the electronic properties of graphene deposited on structures constructed in such way that the properties detected by STS for gold substrate were accurately described. Finally, both simulated and experimental data are compared and assessed to be in a satisfactory agreement.

The paper is organized as follows. In Sec. II the calculation method and experimental details are summarized. In Sec. III we report on the results of STS measurements of the graphene/gold sample and on the dI/dV profiles recorded for pristine surface of gold. The theoretical interpretation of the observed heterogeneity is proposed, analyzed and compared with experimental data in Sec IV. Some final remarks and perspectives are discussed in Sec. V.

\section{Methods}
\subsection{DFT calculations}
First principles calculations based on the DFT have been performed using \textsc{vasp} software\cite{vasp1, vasp2} within the generalized gradient approximation in the parametrization of Perdew, Burke and Ernzerhof (PBE)\cite{pbe} empirically corrected to include long-range dispersion according to a scheme proposed by Grimme \cite{grimme, bucko}. The dispersion coefficient $C_{6}$ for gold has not been included in the original paper of Grimme\cite{grimme}, thus we have used a value of 40.62 Jnm$^{6}$/mol and of 1.772 \AA\, for the vdW radius of Au ($R_{0}$), as proposed in Ref. \onlinecite{adsorption}. The pair interactions up to a radius of 12 \AA\, have been included in the simulations and the global scaling factor $s_{6}$ has been set to 0.75 since the PBE exchange-correlation was employed. Moreover, the projector augmented wave (PAW) pseudopotentials\cite{paw1, paw2} were used for electron-ion interactions and the electronic wave functions have been expanded in a plane-wave basis set of 400 eV. The electronic self-consistency criterion has been set to $10^{-7}$ eV.

\begin{figure}
\includegraphics[width=0.95\columnwidth]{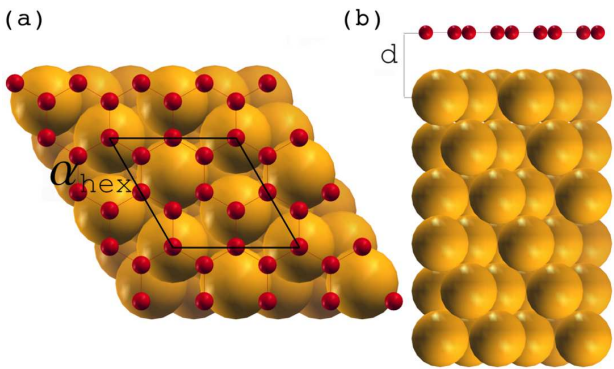}
\caption{\label{geometria} (Color online) Top view (a) and side view (b) of adsorption geometry of graphene on a Au(111). Carbon atoms are denoted as red (darker) balls, gold atoms as yellow (lighter) ones. Parallelogram defines the unit cell.}
\end{figure}

The (111) surface of gold is modeled by a periodic slab geometry. Supercells consisting of six/seven layers of Au atoms and a single sheet of graphene are separated by the vacuum thickness of at least 15 \AA\, placed in the direction normal to the graphene's plane to avoid interactions with spurious images of the slab. We employed the typical in-plane adsorption geometry as shown in Fig.\ref{geometria}, where 2$\times$2 graphene's and substrate's R30$^{\circ}(\sqrt{3}\times\sqrt{3})$ unit cells are directly matched. The size of the supercell is determined by the in-plane lattice parameter \alfa\, denoted in Fig.\ref{geometria}. We have performed calculations only for one value of \alfa\, adapted to the one of graphene which was optimized using DFT-D2 method and estimated to be $4.932$\,\AA\,\cite{nasza6}. The structural relaxations have been performed for the uncovered substrate first, next the graphene sheet has been added above, and all atoms in the top two layers of metal as well as all carbon atoms have been allowed to move. Total energies were converged to within $10^{-6}$ with respect to the ionic steps. The determined equilibrium distance is equal to $d_{\mathrm{eq}}$=3.23\AA.
Other models of graphene/Au interface are considered and described in details in Sec. IV.

The simulations of the LDOS of graphene/Au system have been done according to the scheme proposed in Ref. \onlinecite{nasza_cu}. The typical STS profiles in Tersoff-Hamann approach are obtained by an evaluation of the constant-height charge images for several values of the bias voltage followed by their numerical differentiation. It allows us to calculate the spectrum in the whole space above the crystal. The choice of the particular point above the surface corresponds to the STS data taken under open loop conditions at the fixed position. The spectroscopy simulations for graphene on metallic substrate are usually demanding: at least one k-point every few meV for a specific band is needed\cite{hofer_differential} which means that a very dense k-points mesh must be used in the final calculations of density of states. We use the tetrahedron scheme\cite{tetrahedron} and the $\Gamma$-centered 24$\times$24$\times$1 k-point mesh during the optimization and self-consistent runs for accurate Brillouin zone integrations.

\subsection{Experimental details}
The STM/STS studies of monolayer graphene on polycrystalline gold have been done on the sample prepared by Graphene Industries, UK. The sample consists of graphene flakes transferred on 8 nm Au with 0.5 nm Cr adhesion layer  sputtered onto 100 nm SiO$_{2}$ grown in Si(001). The 8 nm gold layer is thick enough to be stable during graphene flakes deposition and sufficiently thin to be transparent for the visible light.  Thus, it is possible to extend the three-interface Fresnel-law-based model effective for optical graphene identification on widely used air/MG/SiO$_{2}$/Si samples\cite{klusek_ref3} to the five-interface system air/MG/Au/Cr/SiO$_{2}$/Si\cite{klusek_ref4}. This enables the optical identification of graphene flakes after deposition on polycrystalline gold. All the STM/STS experiments have been carried out at room temperature in UHV condition using VT-STM/AFM microscope integrated with the XPS/UPS/AES/LEED/MULTIPROBE P system (Omicron GmbH). The tips have been prepared by mechanical cutting from the 90$\%$Pt-10$\%$Ir alloy wires (Goodfellow) and electrochemical etching of the W wires (Goodfellow). In the STS mode the I/V curves were recorded simultaneously with a constant current image by the use of an interrupted-feedback-loop technique. The obtained data have been used to calculate the first derivative of the tunneling current with respect to the voltage (dI/dV) and to build spatial conductance maps i.e. dI/dV(x,y,E). 

\section{Experiment: results and discussion}
\subsection{STM/STS of monolayer graphene deposited on gold}
\begin{figure}
\includegraphics[width=0.95\columnwidth]{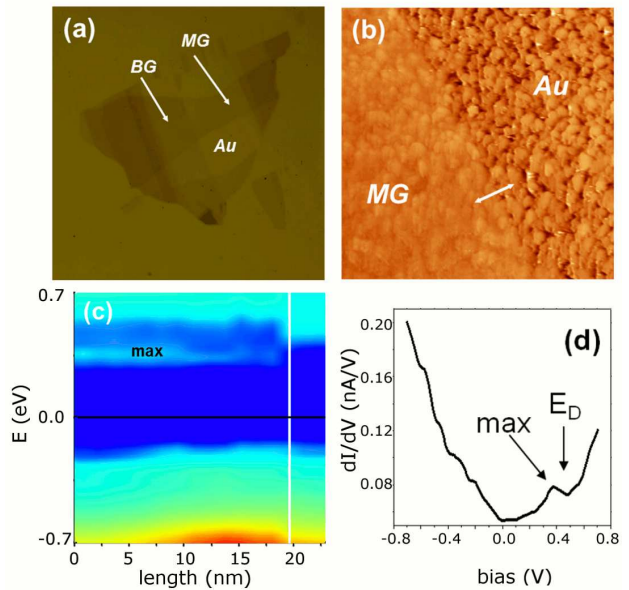}
\caption{\label{probka} (Color online) (a) Optical image of graphene flakes deposited on Au/Cr/SiO$_{2}$/Si substrate. (b) 450 nm x 450 nm STM topography showing the details of MG and Au border line. (c) dI/dV (E, line) map recorded on MG/Au interface along arrow shown in figure (b). Colors: blue, green, red - low, intermediate, and high value of the LDOS, respectively. (d) The example of dI/dV profile recorded on MG/Au sample. }
\end{figure}

The monolayer graphene (MG) deposited on gold substrate has been initially identified by the visual inspection with an optical microscope. As it was mentioned above, the gold layer does not obscure the optical interference effect that makes graphene visible (see Fig.\ref{probka} (a)). Next, the STM topography has been recorded at sample bias U=+0.8 V and with the tunneling current set point equal to 0.2 nA. The 450 nm x 450 nm topography is presented in Fig.\ref{probka}(b) which clearly shows the border between MG/Au and pure Au. It seems that the graphene layer does not map exactly the structure of Au substrate. The estimated value of r.m.s. calculated over 200 nm x 200 nm area on MG equals 0.5 nm while a similar measurement on gold gives value equal to approximately 0.86 nm.

In Fig.\ref{probka} (c) we present dI/dV (E, line) map calculated as a function of the energy (y axis) and of the position along selected line indicated by the arrow in Fig.\ref{probka}(b) (x axis). The most visible feature in the plot is the border between MG/Au and pure Au regions. It is denoted in Fig.\ref{probka}(c) by vertical white line dividing the plot into two parts: the left-hand one shows MG deposited on Au, while the right-hand one is associated with uncovered Au substrate. It is clear that the major change in LDOS starts to appear at MG/Au - Au border. Any strong changes cannot be observed at the occupied part of the spectra, while the dramatic change can be easily noticed in the unoccupied part of the profile, where a considerable decrease in LDOS occurs at about 0.5 eV and is accompanied with the presence of local maximum located at energy equal to 0.37 eV above the Fermi level. The representative dI/dV curve recorded in this region shown in Fig.\ref{probka}(d)  illustrates the asymmetry of the profile with respect to the $E_{F}$. The minimum at 0.5 eV seems to mark the position of the Dirac point ($E_{D}$) of graphene, but only the comparison with DFT results allows for its precise identification\cite{nasza_cu}. Moreover, the energetic position of $E_{D}$ varies across the sample. Taking into account 65536 individual dI/dV curves we estimated that the position of $E_{D}$ is located in the range of 0.25 - 0.55 eV above the Fermi level. 

\subsection{Heterogeneity of graphene's electronic structure}
\begin{figure}
\includegraphics[width=0.95\columnwidth]{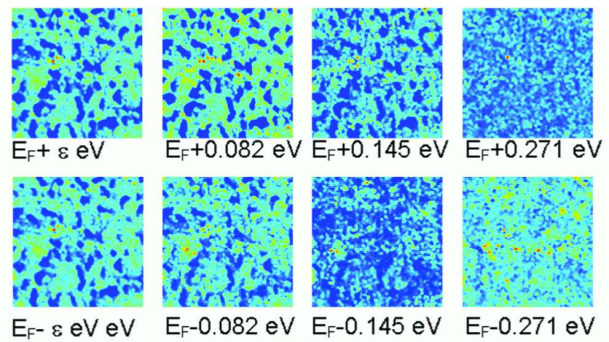}
\caption{\label{hetero} (Color online) 100 nm x 100 nm dI/dV (E, x, y) maps recorded on MG/Au. Colors: blue, green, red - low, intermediate, and high value of the LDOS, respectively.}
\end{figure}

\begin{figure}
\includegraphics[width=0.95\columnwidth]{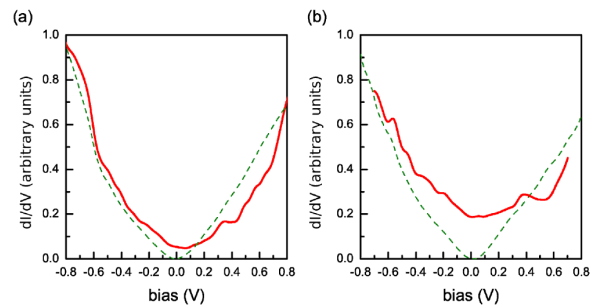}
\caption{\label{spektra} (Color online) (a, b)  dI/dV profiles recorded in different places on MG/Au sample. The dashed green lines correspond to the points in regions denoted as blue in Fig.\ref{hetero}, while the spectra taken in the points within orange/yellow regions are marked as solid red ones.}
\end{figure}

\begin{figure}
\includegraphics[width=0.95\columnwidth]{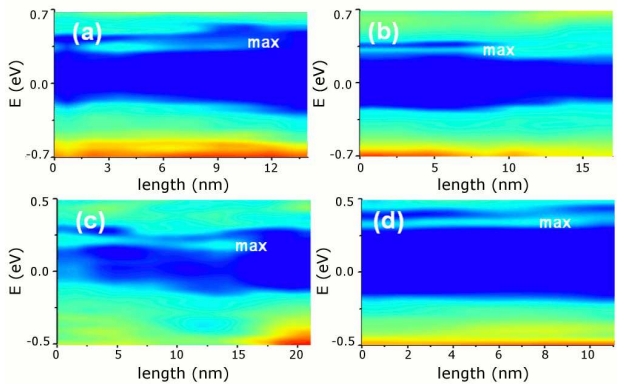}
\caption{\label{linie} (Color online) (a, b, c, d) dI/dV (E, line) maps recorded on different MG/Au regions. Colors: blue, green, red - low, intermediate, and high value of the LDOS, respectively.}
\end{figure}

\begin{figure*}
\includegraphics[scale=1.0]{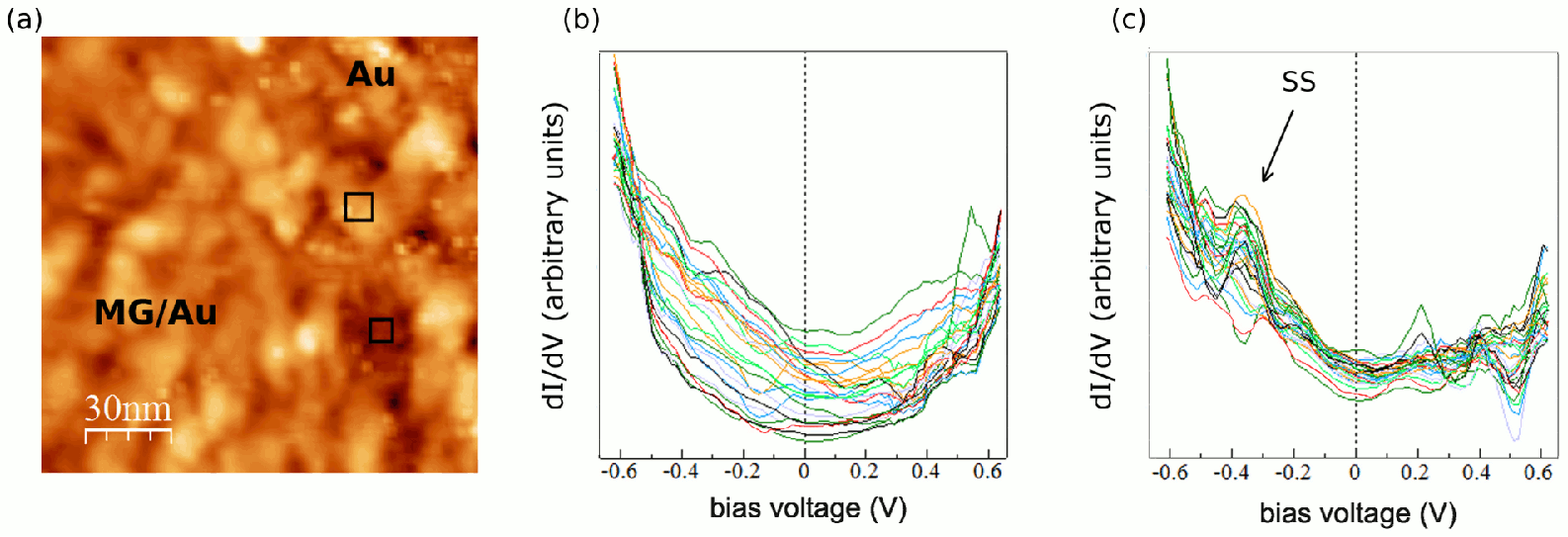}
\caption{\label{pure} (Color online) (a) 150 nm x 150 nm STM topography showing the details of MG and Au border.(b),(c) dI/dV profiles recorded on uncovered Au part of the sample in the regions denoted by squares in figure (a).}
\end{figure*}

The STM topography shown in Fig.\ref{probka} (b) suggests strong stuctural heterogeneity of the substrate, in particular the presence of numerous tiny crystallites of Au, which can be easily observed in both regions, covered and uncovered. Since the grains of Au may differ from each other, it can lead to the diversity in the local electronic properties. We have analyzed the considered MG/Au system using the  conductance maps measured across the sample. In Fig.\ref{hetero} we present 100 nm$\times$100 nm dI/dV (x, y, E) maps at selected energies around the Fermi level. It is clear that the spatial distribution of the LDOS amplitudes recorded at the same energy is not uniform and varies from region to region. It leads to the conclusion that MG is not homogenous in terms of local electronic structure, especially in the case of energies very close to the Fermi level. At energies higher than $E_{F}\pm$0.250 eV spatial distribution of the LDOS seems to be much more uniform and it is difficult to distinguish between different domains. The dI/dV profiles measured in the regions with high LDOS very close to the Fermi level (yellow and light blue patches in Fig. \ref{hetero}) show well defined $E_{D}$ located in the range of 0.25 - 0.55 eV above the Fermi level (see solid red lines in Figs.\ref{spektra}(a),(b)). We have estimated that the size of domains with well defined max/$E_{D}$ feature in conductance maps typically spread over 10-25 nm. The dI/dV curves recorded in the dark blue regions (low LDOS close to $E_{F}$) strongly differs from the profile presented in Fig. \ref{probka} (d), i.e. parabolic shape of dI/dV and the lack of well defined $E_{D}$ point is observed (green dashed lines in Figs.\ref{spektra} (a), (b))). It should also be stressed that, surprisingly, the existence of considerable LDOS heterogeneity close to the Fermi level allows us to easily identify regions on MG/Au/SiO$_{2}$/Si surface with well defined $E_{D}$, even though the Dirac point is located at about 0.25-0.55 eV above $E_{F}$. It can be explained by the fact that the presence of $E_{D}$ is associated with high value of LDOS at the Fermi-level (see red solid lines in Figs. \ref{spektra}(a),(b)). 


It can be easily noticed in Fig. \ref{hetero} that apart from large domains one can distinguish yellow/orange and light blue patches. Figures \ref{spektra} (a), (b) show two different dI/dV profiles (red solid lines) taken in two selected points that correspond to the additional fine structure visible in the maps presented in Fig.\ref{hetero}. These two spectra presented in Fig. \ref{spektra} (red solid lines) differ from each other even in the vicinity of the Fermi level, which is especially pronounced while compared with dI/dV curves (green dashed lines in Fig. \ref{spektra}) taken in regions denoted as dark blue in Fig.\ref{hetero}. The differences include also the variation in the position of $E_{D}$, thus, for completeness, we have recorded series of dI/dV (E,line) maps across the sample (see Fig. \ref{linie} (a)-(d)). Both the intensity and the position of the observed local minimum depends on the spatial position on the investigated surface, which is very prominent on these plots. The signatures that seem to indicate $E_{D}$ positions are placed between 0.25-0.55 eV (compare for example Figs \ref{linie} (a) and (c)), in agreement with preliminary data, but also local variations in electronic structure can be easily noticed (slight deviations from definite values observed along the lines, for example the one in Fig.\ref{linie} (c)). Thus, the existence of large domains may be associated with different properties of visible crystallites, but the influence of other structural factors must also be taken into account. 

It seems that the doping level is determined by the combinations of many factors which can be identified depending on the scale of the observation. For example, the presence of repeated regions with well defined $E_{D}$ above the Fermi-level and those revealing featureless spectra seem to be related mainly to structural properties of different crystallites of the substrate (e.g. amorphous and crystalline  domains), but also to its corrugation as well as to the local variations in the relaxation of the surface region. It seems, however, that in the case of the local changes illustrated in Fig. \ref{linie} rather the influence of last two factors should be dominant. In order to demonstrate this hypothesis, we have studied pure Au substrate to provide information concerning its structural and electronic heterogeneity needed in further DFT simulations.

\subsection{STM/STS of pure Au substrate}
The STM/STS studies of gold have shown strong heterogeneity of electronic properties which seems to be related to the structural diversity as presented in Fig. \ref{pure}(a). We have distinguished two main types of regions different in terms of LDOS - such as those illustrated in Fig. \ref{pure} (b) and (c). Figure \ref{pure} (b) presents rather featureless spectra and reveals asymmetry between occupied and unoccupied states i.e. the LDOS is much higher at occupied states for all energies.  This behavior is rather typical for amorphous gold substrates. In contrast, Fig. \ref{pure} (c) shows spectra with continuous increase of LDOS for occupied states accompanied with clearly visible maxima located close to 0.4 eV below the Fermi level. This behavior seems to be the fingerprint of the Schockley surface state (SS) lying in the projected bulk band gap in $\Gamma$-L direction and observed on the Au(111) surface\cite{klusek_ref1, klusek_ref2}. It leads to the conclusion that the gold substrate is partially covered by the crystallites with (111) orientation.

\section{Theoretical models of graphene/gold interface}
The experimental studies of the sample revealed heterogeneity in the electronic properties of graphene which seems to be domains of p-type and zero doping as well as the slight spread in the values of p-type doping within the respective regions. The STS spectra recorded in certain regions of uncovered gold showed surface states typical for (111) orientation, while featureless characteristics indicating the amorphous metal have been detected in other parts of the sample. Thus, the presence of domains can be associated with the structural diversity of the substrate.

\begin{figure}
\includegraphics[width=0.95\columnwidth]{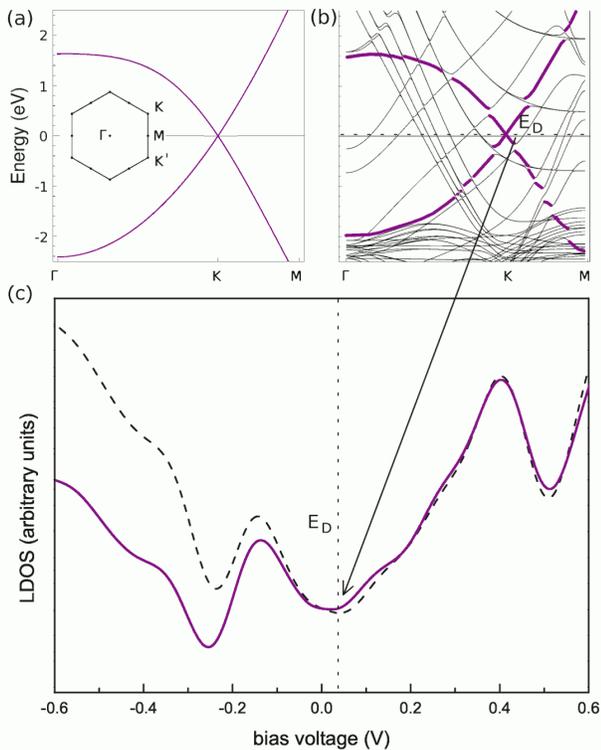}
\caption{\label{zloto111} (Color online) (a) The electronic structure of freestanding graphene in 2$\times$2 unit cell. The inset shows a Brillouin zone with the marked $\Gamma$, K, and M high-symmetry points. (b) The band-structure of graphene on Au(111). (c) The simulated LDOS spectra of a graphene/Au(111) system for different positions above the surface. Fermi level lies at the energy value equal to zero.}
\end{figure}

The interface between the graphene and a fcc metal is usually modeled theoretically as a sheet of graphene lying on a (111) metal surface. The (111) face has been elaborated due to the stability, symmetry and importance of this configuration for practical applications\cite{molekuly1, molekuly2, molekuly3, molekuly4}. First, we have performed the typical DFT simulations for graphene interacting with Au(111). The considered system is presented in Fig. 1, while all the calculations' details were described in Sec. IIA. The calculations with semiempirical van der Waals corrections have shown that the Dirac cones are preserved in the electronic structure, but they are shifted of 0.03 eV above the Fermi level ($\Delta E_{F}=E_{D}-E_{F}=+0.03$ eV). The obtained band-structure is shown in Fig. \ref{zloto111} (b). One can notice that it agrees well with previous theoretical results\cite{holendrzy, nasza6}. It should be also reminded that the level of doping is very sensitive to the distance between the substrate and the graphene sheet\cite{nasza6}. Even its small increase induces higher values of doping.

\begin{figure}
\includegraphics[width=0.95\columnwidth]{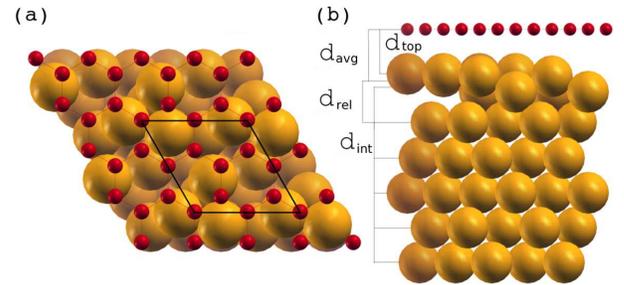}
\caption{\label{amorficzny} (Color online) (a) Top view of adsorption geometry of graphene/Au system with lateral rearrangement of Au atoms in the topmost substrate's layer. (b) Side view of graphene/Au configuration with vertical rearrangements of Au atoms in the topmost substrate's layer. Carbon atoms are denoted as red (darker) balls, gold atoms as yellow (lighter) ones.}
\end{figure}

Figure \ref{zloto111} (c) presents the corresponding LDOS spectra obtained according to the method described in Sec. II A. Two curves  have been simulated for different positions above the surface. Their overall shape is similar to those obtained for graphene interacting with Cu(111) and Pt(111) surfaces\cite{nasza_cu}, but it seems that in the present case the features originating from complicated electronic structure of gold are better pronounced (compare Figs.\ref{pure} (c) and \ref{zloto111} (c)). Even in the LDOS profile calculated just above the surface (dashed line in Fig. \ref{zloto111} (c)) the states of Au(111) can be visible in spite of the fact that carbon states dominate. The position of Dirac point (+0.03 eV) can be associated with the wide global minimum in the profile (dashed line), which is also confirmed by the band-structure plotted for this system (Fig. \ref{zloto111} (b)). These spectra seem to reflect the properties of graphene lying in the regions denoted as blue patches in Fig.\ref{hetero}, whose dI/dV profiles are presented in Figs. \ref{spektra} (a), (b) as green dashed lines. The STS measurements have shown the global minimum localized at the Fermi level which provides a zero value of doping, in satisfactory agreement with theoretical results (+0.03 eV). The simulated profile (solid purple line in Fig.\ref{zloto111} (b)) is quite similar to the recorded one, but it should be noted that the Au(111) surface state is hardly detected by the STS in the regions covered with graphene (see Fig. \ref{spektra}).

\begin{figure}
\includegraphics[width=0.95\columnwidth]{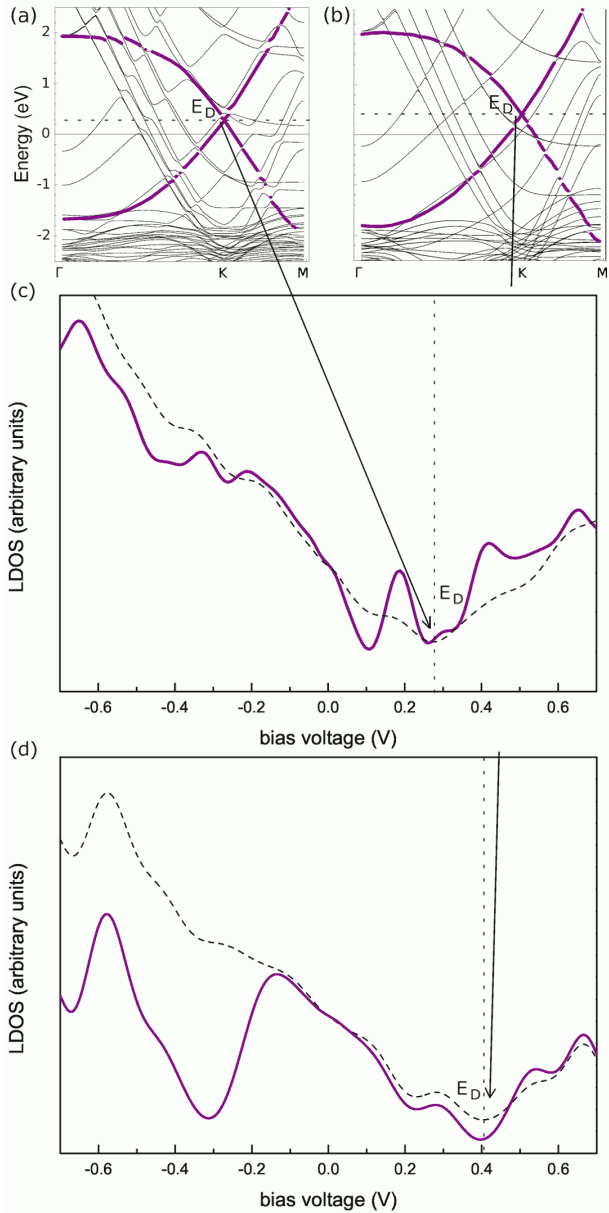}
\caption{\label{amorficzny_bands} (Color online) (a), (b) The band-structures of graphene/Au systems with vertical rearrangements of Au atomic positions. (c),(d) The corresponding simulated LDOS profiles. The zero energy is at a Fermi level.}
\end{figure}

Now, we introduce the interfaces containing strongly rearranged surfaces which can represent amorphous regions in the sample. We define the amorphous-type metallic surface as a set of atoms arranged randomly at the reasonable volume and relaxed. Such choice of the structure is the first step in the method of an \textit{ab initio} random structure searching of the most stable configurations in the given systems\cite{randomstructure}. It has been demonstrated that the relaxations of randomly distributed atoms provide the most stable structure as the one the most frequently obtained among many trials. The other results of optimizations represent possible less stable configurations which should contain also amorphous-like systems. Here, we consider only slight deviations from the (111) arrangement of atoms in a topmost layer of metal. This approach seems to be reasonable due to the extreme sensitivity\cite{nasza6} of the doping level of graphene to any changes in the positions of atoms. The effect of such small variation should be then easily noticed. Moreover, we can distinguish the shifts of atoms with respect to the positions in (111) surface into lateral and vertical directions.

The geometry of the significant lateral rearrangement of atoms in a first layer is shown in Fig.\ref{amorficzny} (a), whereas the slight vertical perturbation is illustrated in Fig.\ref{amorficzny} (b). DFT calculations performed for the first system have shown that the lateral rearrangements do not lead to any variation in adsorption distance and to only a very slight change of the doping level ($\Delta E_{F}$=+0.08 eV) which might be caused by the local increase in distances between particular C-Au pairs (C atoms are not longer on top of Au atoms). The band-structure corresponding to the configuration shown in Fig. \ref{amorficzny} (b) is illustrated in Fig.\ref{amorficzny_bands} (a). One can observe that the conical point is shifted to $\Delta E_{F}=$0.28 eV, a value much larger comparing to results for Au(111) surface (+0.03 eV). The Dirac point position can be also easily recognized in simulated LDOS profile (Fig.\ref{amorficzny_bands} (c)). In this case the shape of the spectrum is more similar to the experimental profiles which seem to correspond to the amorphous regions (solid red lines in Figs.\ref{spektra} (a), (b)). 

The difference in doping between graphene interacting with Au(111) and with amorphous-like surface presented in Fig.\ref{amorficzny} (b) can be explained by the arguments of geometry and symmetry. The gold atoms in the first layer by definition does not lie in the same plane, because their positions are generated randomly in a given region. We can always distinguish the topmost atom, the middle and the lowest one. The distance between graphene and the plane of the topmost atom ($d_{\mathrm{top}}$)  is typically smaller (2.5-3.1\AA) than the one optimized on perfect Au(111) substrate (3.23 \AA), but the distance between graphene and the lowest atom is defined to be smaller than the average one $d_{\mathrm{avg}}$ (3.2-3.7\AA). Such configuration always leads to increase in doping, which could reach over +0.3 eV, but the accurate value strongly depends on the strength (amplitude) of the assumed perturbation.

It should also be stressed that the average distance between the two topmost layers of gold atoms ($d_{\mathrm{rel}}$) usually changes comparing with the bulk layer spacing ($d_{\mathrm{int}}$=2.36\AA) and with the Au(111) configuration (approximately 2.5 \AA, see Table II in Ref.\onlinecite{nasza6}). This effect is equivalent to average contraction or expansion of the layer spacing (the latter case is illustrated in Fig. \ref{amorficzny} (b)), which was demonstrated to strongly influence the doping level\cite{nasza6}. In particular the increase in surface region interlayer distances raises the doping, while the contraction lowers its value (see Fig. 6 in Ref.\onlinecite{nasza6}). It seems to be the explanation of the local modulations in the value of p-type doping observed in the spectra presented in Fig.\ref{linie}, but can enhance also the differences observed between the values in doping in separate regions on the sample (compare Figs. \ref{linie} (a) and (c)).

It is worthwhile to note that the doping level of perturbed system similar to the one presented in Fig.\ref{amorficzny} (b) does not achieve the maximal value of 0.55 eV observed in the experiment. It is due to the reorganization of bands in the vicinity of the K-point of the Brillouin zone which can be easily observed in Fig. \ref{amorficzny_bands} (a). Values of doping higher than +0.3 eV can be achieved for only slightly perturbed systems, but having expanded two topmost layers of the gold. Such system behaves similarily to the one with increased interlayer spacing which was described in details in Ref.\onlinecite{nasza6} and illustrated in Fig.6 therein. In Fig. \ref{amorficzny_bands} (b) and (d) we show, respectively, the calculated band-structure and LDOS profile for such configuration. The doping level achieves 0.41 eV, which is closer to the maximal measured value. 

Finally, we would like to stress the role of the detected corrugation and the fact that the graphene does not map exactly the structure of gold. The configuration with graphene suspended on Au(111) crystallites and above the amorphous regions provides the differences in the graphene-substrate distance in a natural way. Since the doping level is very sensitive to the graphene-substrate distance\cite{nasza6}, all the described variations in the Fermi level shift might be enhanced.

\section{Final remarks and perspectives}
To successfully employ graphene in the electronic devices and to construct systems promising for applications, it is important to understand the graphene-metal interaction. The preliminary theoretical results\cite{nasza6} showed the sensitivity of the doping of graphene to the geometry of the metallic substrate, which reveals potentially high tunability of such systems. Thus, we have used scanning tunneling spectroscopy to study the properties of graphene deposited on conductive heterogenous substrate and performed DFT calculations to explain the obtained results. 

We have observed the domains of zero and p-type doping across the graphene/Au sample, but it should be stressed that the value of p-type doping differs between individual regions as well as varies slightly within each specific domain. Additional STS studies of pure Au substrate suggested that the existence of large domains is related to the presence of both amorphous and crystalline regions with (111) orientation. Our theoretical studies seem to confirm this hypothesis. Moreover, even the changes in Au atomic positions within the topmost layer of the substrate lead to the transition from zero doping typical for perfect Au(111) substrate to the Fermi level shift below the graphene's Dirac points. The distinct configurations of atoms in amorphous-type system provide different values of the p-type doping, in accordance with experimental data (compare Figs. \ref{linie} (a) and (c)). This factor seems to be also responsible for slight variation in the doping level within selected domains in the sample. The observed modulations might be associated with local changes in the average interlayer spacing in metal. Obviously, the domination of every particular factor depends on the scale of the observation, thus the influence of those with the smallest impact can be detected only by studies of the internal electronic structure of the specific domain on the sample.

Finally, we have shown that gold/graphene interface has no negligible influence on the electronic properties of graphene layer. This leads to the conclusion that this influence must be taken into consideration when the real graphene-based nanoelectronics devices are designed and built.

\begin{acknowledgments}
This work is financially supported by Polish Ministry of Science and Higher Education in the frame of Grant No. N~N202~204737. Two of us (J.S., I.W.) acknowledge support from the European Social Fund implemented under the Human Capital Operational Programme (POKL), Project: D-RIM. Part of the numerical calculations reported in this work have been performed at the Interdisciplinary Center for Mathematical and Computational Modelling (ICM) of the University of Warsaw within the Grant No. G47\,-\,8 and with the support of the "HPC Infrastructure for Grand Challenges of Science and Engineering" Project, co-financed by the European Regional Development Fund under the Innovative Economy Operational Programme. Figure 1 was prepared using the \textsc{xcrysden} program. \cite{kokalj}
\end{acknowledgments}
\end{document}